\documentclass[12pt]{article}

\newcommand{\m}{\mathrm}
\newcommand{\be}{\begin{equation}}
\newcommand{\ee}{\end{equation}}
\newcommand{\ba}{\begin{eqnarray}}
\newcommand{\ea}{\end{eqnarray}}

\usepackage{graphicx}
\usepackage{amssymb}
\usepackage{amsmath}
\usepackage[T1]{fontenc} 
\usepackage[ansinew]{inputenc} 
\usepackage[nosort]{cite}
\newcommand{\inbar}{\vrule height1.57ex width.4pt depth0pt}
\newcommand{\SW}{\relax{\hbox{$\ \inbar\kern-.285em{\rm S}$}}}

\begin{document}
\thispagestyle{empty}
\begin{center}

\null \vskip-1truecm \vskip2truecm

{\Large{\bf \textsf{Slow Complexification}}}

{\large{\bf \textsf{}}}

{\large{\bf \textsf{}}}

\vskip1truecm

{\large \textsf{Brett McInnes}}

\vskip1truecm

\textsf{\\  National
  University of Singapore}

\textsf{email: matmcinn@nus.edu.sg}\\

\end{center}
\vskip1truecm \centerline{\textsf{ABSTRACT}} \baselineskip=15pt
\medskip

The fact that AdS black hole interior geometries are time-dependent presents two challenges: first, to holographic duality (the boundary matter tends to equilibrate, often very quickly), and, second, to the idea that wormholes can be traversable (the wormhole geometry is dynamic, and the wormhole is apt to collapse too quickly for traversal to be possible). As is well known, the first puzzle can be addressed by considering the quantum circuit complexity of the strongly coupled boundary matter, which can continue to grow long after equilibrium is established. We show that data from a phenomenological model of the Quark-Gluon Plasma indicate the existence of an upper bound on the rate of increase of the (specific) complexity, in agreement with a simple holographic model. We then point out that, in this model, this upper bound becomes stricter if angular momentum is added to the bulk black hole while fixing the temperature (at any value, so the black hole is \emph{not} near-extremal). We show that the dual phenomenon, a dramatic slowing of the black hole interior dynamics at high specific angular momentum, also occurs. We conjecture that sufficiently slow complexification of the field theories dual to rotating black holes is associated with traversability of the bulk wormhole, when quantum effects are taken into account.

\newpage

\addtocounter{section}{1}
\section* {\large{\textsf{1. Can Complexification be Stopped?}}}
Asymptotically Anti-de Sitter black holes\footnote{We consider only four-dimensional boundary theories, and so \emph{all black holes in this work are five-dimensional}.} play a central role in the gauge-gravity duality \cite{kn:nat}: their exterior spacetimes model the thermal properties of the dual field theory. Paradoxically, however, they present the duality with a very serious challenge, because it is notoriously difficult to establish a convincing holographic account of black hole \emph{interiors} (see for example \cite{kn:oyc,kn:interior,kn:anything3}).

One problem is that there is no timelike Killing vector field in (at least part of) the interior; thus, the interior geometry is normally highly dynamic, certainly just under the event horizon but probably throughout. However, the exterior and the dual system at infinity apparently do cease to evolve once they equilibrate (as we assume happens for ``large'' AdS black holes), and this can happen very quickly. There is apparently a serious conflict with holographic duality here.

To address this problem, Susskind and others \cite{kn:suss1,kn:suss2,kn:suss3,kn:suss4} proposed the existence of some kind of relation between the dynamics of the black hole interior geometry and the growth of the \emph{quantum circuit complexity} in the dual field theory: the point being that the latter normally continues to grow, as unitary evolution proceeds, long after thermodynamic equilibrium has been established for the boundary matter. (See \cite{kn:poli} for a recent review. Note that work has begun to extend these ideas to the de Sitter case: see \cite{kn:switch} and its references.)

The precise definition of ``complexity'' in this context remains a matter of debate (for example, \cite{kn:anything1,kn:anything2}), and so does the precise relation between boundary complexity and the dynamics of black hole interiors \cite{kn:engel}. Fortunately, one has some general ideas regarding the \emph{rate} at which the complexity, however it is defined, increases. We will focus on the late-time regime in which thermodynamic equilibrium has been established (so that in particular the mass of the black hole is constant), and in which complexity increases at a rate proportional to the entropy and to the temperature (see \cite{kn:tallarita} for a clear discussion). This continues for a time exponential in the entropy of the system \cite{kn:suss5}. That is, we avoid the ultra-late regime, where the complexity ceases to increase, which is not relevant to the situations we will discuss.

Studies of this regime using the holographic duality with the black hole exterior spacetime have led \cite{kn:suss3} to predictions of \emph{upper bounds} on the rate of ``complexification'' of the dual field theory. Such bounds probe the fundamental physics of the boundary matter, and they will be our principal concern here.

Now the very existence of upper bounds on complexity growth has in fact been strongly challenged. At first it was thought that the bounds are due or at least related to the well-known Lloyd bound \cite{kn:lloyd}, but it is now known that this is not correct \cite{kn:jordan,kn:cottrell}, and that there are in fact many systems with unbounded rates of complexification.

It is still possible, however, that although such bounds cannot be established in general, they might still hold for \emph{the kinds of systems that are dual to black holes}, namely systems consisting of strongly coupled matter. In view of the results of \cite{kn:jordan,kn:cottrell}, one is however entitled to be sceptical regarding this. So before we pursue this question into more extreme domains, we should answer a simple question: is there any evidence that complexity growth is actually bounded for \emph{real} strongly coupled matter?

At this point we need a brief detour to argue for a refinement of the concept of complexity growth, however one defines complexity itself. The rate of complexity growth, as it is usually computed, is an extensive quantity. While this may be of interest for some purposes, we wish to argue that it is not as interesting as the rate of complexity growth \emph{allowing for the resources available}. We propose to do this by making the usual move in thermodynamics, where one usually considers quantities defined per unit of mass: in particular, one normally deals with the \emph{specific} entropy, the entropy per unit mass (or per particle in the relativistic case), at a definite temperature. For example, the NIST tables \cite{kn:nist} give molar entropies, which are equivalent to specific entropies, at prescribed temperatures.

In this spirit, we propose to focus on the rate of growth of the ``\emph{specific complexity}'', which we denote by $\mathfrak{C}$, also at fixed temperature. Again, the point is that the rate of growth of $\mathfrak{C}$ at some temperature is being proposed as a fundamental, \emph{intrinsic} property of strongly coupled matter, like the density at a given temperature, a property independent of the size of a particular sample (though, like density, it might depend on other parameters as well).

Resuming our discussion of bounds on rates of complexity growth: the form of strongly coupled matter to which we have most direct access is the Quark-Gluon Plasma (QGP) produced in heavy-ion collisions \cite{kn:franc,kn:solov}. Sophisticated phenomenological models predicting the values of the physical parameters of such plasmas have been developed (a very clear and explicit example is \cite{kn:sahoo}). We can use such models to compute (an approximation to) the rate of specific complexification of a real strongly coupled system, to try to determine whether holographic bounds on such rates are in any sense physical. Surely this is the first question to ask regarding these bounds: after all, if real strongly coupled matter has a rate of (specific) complexification which shows no sign of being bounded with respect to the variations of some well-understood physical parameters, then the whole question is of doubtful interest.

For the QGP, the principal parameters are the impact energy of the collision that produces it, and the impact parameter of that collision. The former is correlated with the initial \emph{temperature} of the plasma, while the latter controls the \emph{specific angular momentum} (the ratio of the angular momentum and energy densities; again, this is much more useful than the total angular momentum); see \cite{kn:becca} and \cite{kn:captain} for recent discussions, with further references. The holographic dual system is a ``large'' AdS$_5$ black hole with a non-zero temperature and specific angular momentum: that is, a (non-extremal) AdS$_5$-Kerr black hole.

We will see that this holographic model predicts the existence of a universal upper bound on the rate of increase of the specific complexity of the boundary theory, as it varies with temperature and specific angular momentum. This bound is on the order of $10^{51}\,/(\m{kg}\cdot \m{s})$. We will also see that the phenomenological data indicate that, huge though this rate may be for ordinary matter, it is in fact nearly attained by the QGP produced in low-centrality heavy ion collisions. According to the holographic model, then, these plasmas are the \emph{fastest} ``complexifiers''.

The comparison we have been discussing is at low (specific) angular momenta, because our data are taken from \cite{kn:sahoo}, which is concerned with idealised nearly central collisions. The key question now is this: if we study peripheral (non-central) collisions producing plasmas with higher specific angular momenta, will the holographic upper bound on the rate of specific complexification continue to hold?

The answer is affirmative: this is an immediate consequence of the findings of \cite{kn:109}, where we found that (with minor exceptions of no real phenomenological importance) rotation tends to \emph{decrease} the specific entropy of a five-dimensional asymptotically AdS black hole when the comparison is made at fixed Hawking temperature, and this does indeed suppress the specific complexification rate of the dual system; so the upper bound found in the slow-rotation case continues to hold.

The crucial question now is: what happens when the specific angular momentum continues to increase? We find that the holographic model makes a very surprising prediction: the rate of specific complexification of the boundary matter can be driven down, at fixed temperature, to \emph{arbitrarily small} values by taking the specific angular momentum (of either the bulk black hole or the boundary matter) to be sufficiently large. If this effect is real, then, by taking the specific angular momentum of the plasma to be sufficiently large, we can force this system to be the \emph{slowest} ``complexifier''.

According to the holographic understanding of this situation, a strong suppression of complexification by large specific angular momenta should be reflected in a similarly large suppression of the rate at which the bulk black hole interior evolves.

We will show explicitly that, at least in the region of the interior which can be described approximately by the standard metric, this is in fact the case. (This is not straightforward, because the relation between angular momentum and spacetime geometry is very much more complex in the AdS$_5$ case than for four-dimensional asymptotically flat Kerr geometries.)

One of the most interesting aspects of the time-dependence of black hole interiors is the phenomenon of the collapse of the ``wormhole'' which exists in the idealised case. Our results imply that large specific angular momenta delay this collapse, and it is natural to ask whether this is connected with the much-discussed question of \emph{traversability} \cite{kn:aronwall,kn:juan} for wormholes in rotating black holes. Indeed, it has recently been shown that backreaction effects can induce traversability of wormholes in such black holes \cite{kn:bilotta} (though this is in the four-dimensional, asymptotically flat context). We conjecture that, in the AdS$_5$-Kerr case, the phenomenon of ultra-slow complexification of boundary matter is the holographic dual of the traversability of quantum-perturbed wormholes in black holes with high specific angular momenta. We conclude with a speculation on the relation of complexity to quantum teleportation, which is thought \cite{kn:aronwall,kn:juan} to be dual to bulk traversablity.

We note in passing that the idea of relating the physics of AdS black hole interiors to that of the Quark-Gluon Plasma is not new; it has recently been discussed in impressive detail in \cite{kn:andreas}. It should also be noted that the complexity of systems dual to rotating black holes was investigated (from a very different point of view) in \cite{kn:robie}.

We begin with a brief review of the relevant aspects of the AdS$_5$-Kerr spacetime.

\addtocounter{section}{1}
\section* {\large{\textsf{2. Geometry and Physics of AdS$_5$-Kerr}}}
The metric for the AdS$_5$-Kerr black hole (in the case of rotation around a single axis, the only case we consider\footnote{The metric in the case where the black hole rotates simultaneously around two axes is also known explicitly, but we have no need of this additional complication.}) is given by \cite{kn:hawk,kn:cognola,kn:gibperry}
\begin{flalign}\label{A}
g(\textsf{AdSK}_5)\; = \; &- {\Delta_r \over \rho^2}\left[\,\m{d}t \; - \; {a \over \Xi}\,\m{sin}^2\theta \,\m{d}\phi\right]^2\;+\;{\rho^2 \over \Delta_r}\m{d}r^2\;+\;{\rho^2 \over \Delta_{\theta}}\m{d}\theta^2 \\ \notag \,\,\,\,&+\;{\m{sin}^2\theta \,\Delta_{\theta} \over \rho^2}\left[a\,\m{d}t \; - \;{r^2\,+\,a^2 \over \Xi}\,\m{d}\phi\right]^2 \;+\;r^2\cos^2\theta \,\m{d}\psi^2 ,
\end{flalign}
where
\begin{eqnarray}\label{B}
\rho^2& = & r^2\;+\;a^2\cos^2\theta, \nonumber\\
\Delta_r & = & \left(r^2+a^2\right)\left(1 + {r^2\over L^2}\right) - 2M,\nonumber\\
\Delta_{\theta}& = & 1 - {a^2\over L^2} \, \cos^2\theta, \nonumber\\
\Xi & = & 1 - {a^2\over L^2}.
\end{eqnarray}
Here $L$ is the background AdS curvature length scale, and $M$ and $a$ are geometric parameters which are related only remotely to the physical mass and the angular momentum per unit mass (see below). The angular coordinates on the (topological) $r = $ constant three-spheres are such that $\phi$ and $\psi$ run from $0$ to $2\pi,$ but $\theta$ runs from $0$ to $\pi/2$ (not $\pi$). This metric defines a conformal geometry at infinity in the usual manner: the spatial geometry is that of a rotating spheroid.

It is immediately clear from the form of this metric that we must have either $a/L < 1$ or $a/L > 1.$ The latter case is important theoretically \cite{kn:104}, but it does not arise in the application to strongly coupled matter. Henceforth, then, we enforce $a/L < 1,$ the inequality being strict.

We now discuss the physical parameters of this black hole. For reasons to be explained, we use (until further notice) conventional rather than natural or Planck units.

First, the entropy of the black hole is (from (\ref{A}); notice the factors of $1/\Xi$ in the terms involving $\phi$)
\begin{equation}\label{C}
S_{\textsf{BH}}\; =\; {\pi^2k_{\textsf{B}}c^3\left(r_{\textsf{H}}^2 + a^2\right)r_{\textsf{H}}\over 2G_5 \hbar\,\left(1 - {a^2\over L^2}\right)},
\end{equation}
where $k_{\textsf{B}}$ is the Boltzmann constant, $c$ is the speed of light, $G_5$ is the five-dimensional gravitational constant, $\hbar$ is the reduced Planck constant, and $r_{\textsf{H}}$ is the value of the radial coordinate at the event horizon (assumed, throughout this work, to exist; see in this connection \cite{kn:dad}).

At this point, we draw the reader's attention to two simple observations. The first is that the entropy depends explicitly on $G_5$. If we attempt to move to the boundary, we find \cite{kn:nat} that $G_5$ does have an interpretation there, but only in terms of a quantity (the number of colours) which is not actually known. This is a problem if we wish to use explicit numerical data, as we hope to do here.

The second and more serious point is that the \emph{total} entropy of the black hole is not a useful quantity when we try to transfer it to the boundary. In studies of the QGP, one only speaks of either the entropy \emph{density} or, as in most applications of thermodynamics, of the \emph{specific} entropy (entropy per unit of mass/energy; that is, the ratio of the entropy and energy densities). This is the same argument we made in the preceding Section in favour of a focus on the specific complexity. The holographic dual is the specific entropy (entropy divided by mass) of the black hole, which we now compute.

The physical mass (the one that appears in the First Law of thermodynamics) is given \cite{kn:hawk,kn:cognola,kn:gibperry} by\footnote{A modification of the standard expression for the physical mass of dS and AdS black holes has been proposed, from two different points of view, in \cite{kn:piotr} and \cite{kn:gaogao}. Adopting this modification has interesting consequences but does not alter the conclusions of this work. We will return to this elsewhere.}
\begin{equation}\label{D}
\mathcal{M}_{\textsf{BH}}\;=\;{\pi M c^2\left(3 - {a^2\over L^2}\right)\over 4\,G_5\,\left(1 - {a^2\over L^2}\right)^2}.
\end{equation}
The specific entropy of the black hole is therefore
\begin{equation}\label{E}
\mathfrak{s}_{\textsf{BH}}\;=\;{4 \pi k_{\textsf{B}} c r_{\textsf{H}}\left(1 - {a^2\over L^2}\right)\over \hbar \left(3 - {a^2\over L^2}\right)\left(1 + {r_{\textsf{H}}^2\over L^2}\right)}.
\end{equation}
We notice at once that this does \emph{not} involve $G_5$; so the physically well-motivated replacement of the entropy by the specific entropy resolves this problem.

Precisely the same observation applies to the angular momentum: the quantity of interest for the QGP is not the total angular momentum, but rather the specific angular momentum (the ratio of the angular momentum and energy densities). The angular momentum of the black hole is
\begin{equation}\label{F}
\mathcal{J}_{\textsf{BH}}\;=\;{\pi M c^3 a\over 2\,G_5\,\left(1 - {a^2\over L^2}\right)^2},
\end{equation}
which, again, involves $G_5$; but the specific angular momentum, $\mathfrak{j}_{\textsf{BH}}$,
\begin{equation}\label{G}
\mathfrak{j}_{\textsf{BH}}\;=\;{2 a c \over 3 - {a^2\over L^2}},
\end{equation}
does not.

Notice that the parameter $a$ has no direct physical meaning; but equation (\ref{G}) can be solved for it as a function of the actual physical parameter, $\mathfrak{j}_{\textsf{BH}}.$ Henceforth we regard $a$ as this function.

Notice too that our earlier restriction, $a/L < 1,$ can easily be shown to imply (and to be implied by) $\mathfrak{j}_{\textsf{BH}}/L < c.$ We propose to interpret this holographically in the simplest way: when $\mathfrak{j}_{\textsf{BH}}/L$ is close to $c$, this means that the vorticity of the boundary matter is so large that parts of it are moving at speeds close to that of light. We will discuss this further, below.

The requirement that $\mathfrak{j}_{\textsf{BH}}/L$ should be bounded by $c$ has the following crucially important consequence: there are \emph{two} distinct senses in which an AdS$_5$-Kerr black hole may be said to rotate ``rapidly''. The first is the one familiar from the asymptotically flat case: ``rapid'' rotation means that the black hole is close to being extremal, which in physical terms means that its Hawking temperature is extremely small. But now we see that ``rapid'' rotation can also mean that the specific angular momentum, scaled by $L$, is close to its maximal possible value, $c$. The two senses are independent, because one can fix the Hawking temperature at some large value (as indeed is necessary when we discuss the QGP) and still increase $\mathfrak{j}_{\textsf{BH}}/L$ upwards until it nearly reaches $c$. This new definition of ``rapid'' rotation is the physically relevant one in this case, and it is the one we will use exclusively in this work: our black holes are never extremal or indeed near-extremal\footnote{This means, in particular, that the arbitrarily small rates of complexification we will later discover cannot be prevented by means of an appeal to Cosmic Censorship or to the Weak Gravity Conjecture \cite{kn:motl,kn:kats}.}.

We should mention in this regard recent work \cite{kn:horo1,kn:horo2,kn:108,kn:horo3} which very strongly suggests that (topologically spherical) near-extremal black holes are pathological, and that the description of black holes which rotate rapidly in \emph{this} sense differs greatly from the standard Kerr and AdS-Kerr metrics. (See also \cite{kn:tur}.) Again, we avoid these difficulties simply by staying far away from the near-extremal cases.

The Hawking temperature of this black hole is given by
\begin{equation}\label{H}
T_{\textsf{BH}}\;=\;{\hbar c r_{\textsf{H}}\over 2 \pi k_{\textsf{B}}}\left[{1 + {r_{\textsf{H}}^2\over L^2}\over r_{\textsf{H}}^2 + a^2} \,+\, {1\over L^2}\right].
\end{equation}
This, too, does not depend on $G_5$ (unless one wants to express the temperature in terms of the mass, which is \emph{not} our intention here). Transferring it to the boundary, where of course it corresponds to the temperature of the boundary matter, is therefore straightforward.

Equation (\ref{H}) can be solved for $r_{\textsf{H}}$, which we therefore henceforth regard as a known function of $T_{\textsf{BH}}$ and $a$ (therefore, of $T_{\textsf{BH}}$ and $\mathfrak{j}_{\textsf{BH}}$). \emph{Then equation (\ref{E}) allows us to think of the specific entropy $\mathfrak{s}_{\textsf{BH}}$ as a function of $T_{\textsf{BH}}$ and $\mathfrak{j}_{\textsf{BH}}$.} We will use this construction to compute the rate of specific complexification in terms of the physical parameters $T_{\textsf{BH}}$ and $\mathfrak{j}_{\textsf{BH}}$.

The geometric parameters determining the metric, $a$ and $M$, can be regarded in the same way, by using equation (\ref{G}) (to express $a$ in terms of $\mathfrak{j}_{\textsf{BH}}$) or the definition of $r_{\textsf{H}}$, namely
\begin{equation}\label{I}
M\;=\; {1\over 2}\left(r_{\textsf{H}}^2+a^2\right)\left(1 + {r_{\textsf{H}}^2\over L^2}\right),
\end{equation}
to express $M$ in terms of $a$ and $r_{\textsf{H}}$ (that is, in terms of $T_{\textsf{BH}}$ and $\mathfrak{j}_{\textsf{BH}}$).

We are interested in the case in which the black hole (and therefore the boundary system) behaves well from a thermodynamic point of view, by which we mean that it can come into equilibrium with its Hawking radiation \cite{kn:ruong}. This is only possible if the temperature is greater than or equal to $\sqrt{2}\hbar c / \left(\pi k_{\textsf{B}}L\right)$, which we assume to hold henceforth\footnote{Later we will see, in the context of the RHIC collisions, that $L$ is at least 72 femtometres, so this temperature is no greater than about 1.2 MeV, far below the temperature of any quark-gluon plasma; hence this condition is always satisfied in that context.}. This is the point at which we explicitly exclude extremal black holes.

In this case, there is more than one possible value for $r_{\textsf{H}}$, for fixed $T_{\textsf{BH}}$ and $\mathfrak{j}_{\textsf{BH}}$. We always choose the largest value, corresponding to a so-called ``large'' black hole, since these are the black holes which behave well in the above sense. They also have the virtue of always being stable against superradiance \cite{kn:109}.

It is interesting to ask what becomes of the entropy, physical mass $\mathcal{M}$, and physical angular momentum $\mathcal{J}$ when $\mathfrak{j}_{\textsf{BH}}/L$ approaches its upper bound, $c$. As they depend on $r_{\textsf{H}}$ and $M$, which are themselves functions of $T_{\textsf{BH}}$ and $\mathfrak{j}_{\textsf{BH}}$ (that is, their values are not fixed in this discussion), this is far from obvious; but we can settle it as follows.

For ``large'' black holes, $r_{\textsf{H}}$ is, for each fixed value of the temperature, an \emph{increasing} function \cite{kn:109} of $\mathfrak{j}_{\textsf{BH}}$. However, it is bounded above, and in fact it approaches $\pi k_{\textsf{B}} T_{\textsf{BH}} L^2/ \left(\hbar c\right)$ as $\mathfrak{j}_{\textsf{BH}}/L$ tends to $c$. Therefore, according to equation (\ref{I}), $M$ likewise increases with $\mathfrak{j}_{\textsf{BH}}$, but is similarly bounded above, by ${1\over 2}L^2 \left(1 + {\pi^2 k^2_{\textsf{B}} T^2_{\textsf{BH}}L^2\over \hbar^2 c^2}\right)^2$.

In view of these observations, one can now deduce from equations (\ref{C}), (\ref{D}), and (\ref{F}) that, if the temperature is fixed, the entropy, the physical mass\footnote{The function ${3 - {a^2\over L^2}\over \left(1 - {a^2\over L^2}\right)^2}$ is monotonically increasing (in $a$, therefore also in $\mathfrak{j}_{\textsf{BH}}$) on the physical domain, $0 \leq a/L < 1$.}, and the angular momentum (and therefore, in the holographic picture, the entropy, energy, and angular momentum densities of the strongly coupled matter on the boundary) all increase as $\mathfrak{j}_{\textsf{BH}}/L$ increases towards $c$. However, the behaviour of the \emph{specific} entropy (equation (\ref{E}) cannot be determined by inspection (except when $\mathfrak{j}_{\textsf{BH}}/L$ is very close to $c$): a detailed investigation \cite{kn:109} is required, and it shows that, with a few minor exceptions when the temperature is very low (that is, comparable to $\sqrt{2}\hbar c / \left(\pi k_{\textsf{B}}L\right)$), this crucial quantity is actually a \emph{decreasing} function of the specific angular momentum. This will be important later.

We have argued that complexification is of greatest interest when we take into account the resources available, which we do by concentrating on the \emph{specific} complexity, $\mathcal{C}$. But we have just seen that there are also cogent technical reasons arguing in favour of doing this: by using the specific entropy, specific angular momentum, and specific complexity, we eliminate parameters like $G_5$ which are not known to us. (Incidentally, a further reason for thinking the specific complexity a natural object to consider is that its rate of change is dimensionless in natural units, a fact which will be useful later.)

\addtocounter{section}{1}
\section* {\large{\textsf{3. Rate of Change of the Specific Complexity: Small $\mathfrak{j}_{\textsf{BH}}$}}}
Let $\mathfrak{s}_{\partial}$ be the specific entropy of the boundary system (that is, of a form of matter dual to an AdS black hole). Then $\mathfrak{s}_{\partial}/k_B$ can be regarded as a measure of the number of degrees of freedom per particle available to ``execute gates'' \cite{kn:suss2}. We will take this to mean \cite{kn:tallarita} that the number of degrees of freedom relevant here is a dimensionless multiple, of order unity, of $\mathfrak{s}_{\partial}/k_B$.

To evaluate the rate at which these gates are executed, one needs a time scale, and this is naturally supplied by $\hbar /\left(k_BT_{\partial}^{\#}\right)$, where $T_{\partial}^{\#}$ is the temperature of the boundary matter measured in a frame that rotates with that matter. Thus the rate at which gates are executed per particle, measured by an observer who co-rotates ---$\,$ that is, the rate of growth of the specific complexity of the boundary matter according to this observer\footnote{In \cite{kn:suss3}, another expression, of the form $\int TdS$, is proposed, but it is explained there that this differs significantly from the expression used here and in \cite{kn:tallarita} only in the near-extremal case; which, for reasons explained earlier, we avoid throughout this work.} ---$\,$ is given by $\mathfrak{s}_{\partial}T_{\partial}^{\#}/\hbar$. (This argument does not apply at ultra-late times \cite{kn:tallarita}, but this will not be relevant here.)

However, this rate is measured using a co-rotating time coordinate $t^{\#}$, which differs from the time $t$ measured by a non-rotating observer at infinity; this $t$ coincides with the time coordinate in the bulk, as used in equation (\ref{A}). On the other hand $T_{\partial}^{\#}$ also differs from the temperature $T_{\partial}$ measured by that observer. As is explained in \cite{kn:jacob}, the two effects cancel: $T_{\partial}^{\#}dt^{\#} = T_{\partial}dt$. Thus the rate of specific complexification observed by the non-rotating observer is $\mathfrak{s}_{\partial}T_{\partial}/\hbar$.

We propose that $\mathfrak{s}_{\partial}$ corresponds holographically to the specific entropy $\mathfrak{s}_{\textsf{BH}}$ of the bulk black hole, just as $T_{\partial}$ corresponds to the Hawking temperature in the bulk. This is of course a \emph{proposal}; one would certainly like to see how this works in detail. Let us proceed on the basis of this assumption.

We have now, up to a dimensionless factor of order unity,
\begin{equation}\label{J}
{d\mathfrak{C}\over dt} \; = \; \mathfrak{s}_{\textsf{BH}}T_{\textsf{BH}}/\hbar,
\end{equation}
for the kind of matter that is dual to an AdS black hole. Similarly, we assume that the ratio of the boundary theory angular momentum density to its energy density is dual to the black hole angular momentum parameter $\mathfrak{j}_{\textsf{BH}}$.

Substituting equation (\ref{E}) into this equation, we obtain
\begin{equation}\label{K}
{d\mathfrak{C}\over dt} \; = \; {4 \pi k_{\textsf{B}} c T_{\textsf{BH}} r_{\textsf{H}}\left(1 - {a^2\over L^2}\right)\over \hbar^2 \left(3 - {a^2\over L^2}\right)\left(1 + {r_{\textsf{H}}^2\over L^2}\right)}.
\end{equation}

We saw earlier that $r_{\textsf{H}}$ and $a$ can be regarded as functions of $T_{\textsf{BH}}$ and $\mathfrak{j}_{\textsf{BH}}$. It follows that (\ref{K}) gives us the rate of specific complexification in terms of prescribed values of $T_{\textsf{BH}}$ and $\mathfrak{j}_{\textsf{BH}}$. (Henceforth we drop the subscripts, letting the context determine whether we are referring to quantities defined in the bulk or on the boundary.) We can therefore write the rate of specific complexification in the form
\begin{equation}\label{L}
{d\mathfrak{C}\over dt} \; = \; {c^2\over \hbar}\,\Gamma\left(T^*L,\;\mathfrak{j}/(cL)\right),
\end{equation}
where for typographical convenience we define $T^* = k_{\textsf{B}}T/(\hbar c)$ (so that $T^*L$ is dimensionless, like $\mathfrak{j}/(cL)$), and where
\begin{equation}\label{M}
\Gamma\left(T^*L,\;\mathfrak{j}/(cL)\right) \;=\; {4 \pi T^* r_{\textsf{H}}\left(1 - {a^2\over L^2}\right)\over \left(3 - {a^2\over L^2}\right)\left(1 + {r_{\textsf{H}}^2\over L^2}\right)}
\end{equation}
is dimensionless (both in conventional and in natural\footnote{By ``natural'' units, we mean units in which $k_{\textsf{B}},$ $c$, and $\hbar$ are equal to unity, and the unit of energy is the MeV (and the reciprocal length and time unit is the femtometre), as is customary in the heavy ion collision literature.} units).

Unfortunately, $\Gamma\left(T^*L,\;\mathfrak{j}/(cL)\right)$ (obtained, as we recall, by using (\ref{G}) to express $a$ in terms of $\mathfrak{j},$ and then using (\ref{H}) to express $r_{\textsf{H}}$ in terms of $T$ and $\mathfrak{j}$) is surprisingly and inordinately complicated. It can be written out explicitly, but not usefully.

However, we can understand this function by fixing one of the parameters, and studying $\Gamma\left(T^*L,\;\mathfrak{j}/(cL)\right)$ as a function of the other. We begin by fixing $\mathfrak{j}$; thus $\Gamma$ becomes a function of the temperature.

One finds in all cases that $\Gamma\left(T^*L,\;\mathfrak{j}/(cL)\right)$, as a function of $T^*L$, is a monotonically increasing but \emph{bounded} function. The graph of this function (which coincides with $\mathfrak{s}T$ when natural units are used) in the case of a small (in a sense to be explained) value of the specific angular momentum, $\mathfrak{j}/(cL) = 0.2,$ is shown in Figure 1.
\begin{figure}[!h]
\centering
\includegraphics[width=0.95\textwidth]{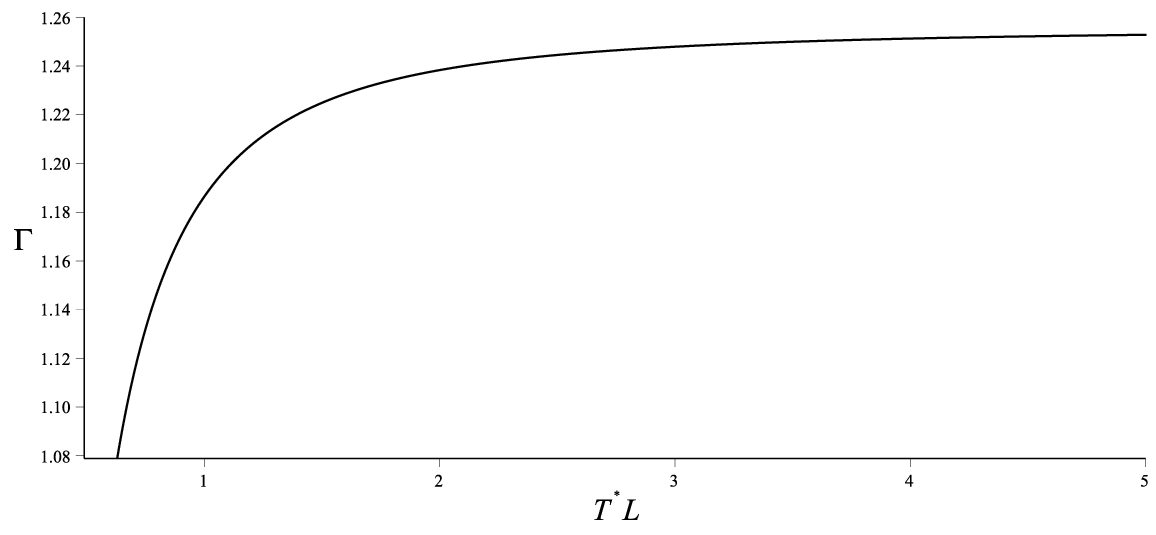}
\caption{ $\Gamma\left(T^*L,\;\mathfrak{j}/(cL)\right)$ for AdS$_5$-Kerr as a function of $T^*L$ when the specific angular momentum is small, $\mathfrak{j}/(cL) = 0.2$.}
\end{figure}

For some perspective, let us actually evaluate the bound in conventional units. Then (since $\Gamma\left(T^*L,\;\mathfrak{j}/(cL)\right)$ is dimensionless also in these units) the upper bound is close to $1.25 c^2/\hbar \approx 1.07 \times 10^{51}/(\m{kg}\cdot \m{s}).$ By contrast, the specific entropy of water \cite{kn:nist} at 300 K is approximately  $3890 \, \m{J}/(\m{K} \cdot \m{kg})$, so the comparable quantity $\mathfrak{s}T/\hbar$ for water is about $1.11 \times 10^{40}/(\m{kg}\cdot \m{s}),$ about 11 orders of magnitude smaller than our bound. In short, for any ordinary form of matter, $\mathfrak{s}T$ would be invisibly small if we tried to plot its values on a graph with the scale used in Figure 1.

Of course, the QGP is not an ``ordinary form of matter''. But it remains the case that we have no reason to expect that the values of $\mathfrak{s}T$ for the QGP will be visible if we plot them on this same scale: they could easily be far too small ---$\,$ or far too large. Furthermore, there is no reason to expect them to show any sign of being bounded.

We propose to study this question using the phenomenological model described in \cite{kn:sahoo}, which is designed to deal with the initial properties (temperatures, energy densities, entropy densities, and so on) of plasmas produced in approximately central (up to 10$\%$ centrality) heavy-ion collisions at the impact energies studied by the STAR collaboration \cite{kn:STAR} at the RHIC facility. We obtain\footnote{Reference \cite{kn:sahoo} reports (Table 1) the values of $s/T^3$ (where $s$ is the entropy density) and $\varepsilon/T^4$ (where $\varepsilon$ is the energy density). The ratio of these two dimensionless (in natural units) quantities is $\mathfrak{s}T$.} in this way Figure 2.

\begin{figure}[!h]
\centering
\includegraphics[width=0.99\textwidth]{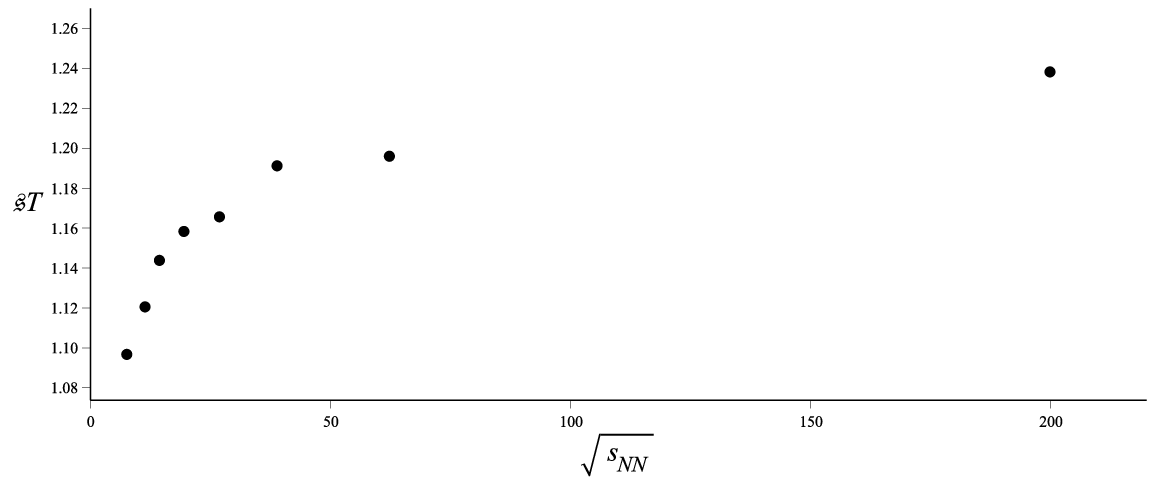}
\caption{The quantity $\mathfrak{s}T$ (which is dimensionless in natural units), numerical data from \cite{kn:sahoo}. The horizontal axis is the impact energy per pair, in GeV.}
\end{figure}

We wish to emphasise that the horizontal axes in Figures 1 and 2 involve two correlated but quite different quantities (temperature and impact energy), and, more importantly, that \emph{we do not know} the numerical value of $L$; we only have a lower bound for it, as mentioned earlier\footnote{In principle, though not in practice, the numerical value of $L$ can be computed holographically by means of the relation ${\ell_s^4\over L^4}\;=\;{1\over \lambda},$ where $\ell_s$ is the string length scale, and where $\lambda$ is the 't Hooft coupling parameter of the boundary field theory.}. Thus a comparison cannot be direct, and cannot be very precise (which is why we have not reported error bars, though, as can be seen from \cite{kn:sahoo}, these would not in fact be very large). We suggest that the reader interpret Figure 1 simply as a demonstration that the holographic model does indeed predict the existence of an upper bound, and that it predicts the magnitude of that bound.

Even with these precautions, however, two things are clear. First, the range of the dimensionless quantities on the vertical axis in Figure 2 is quite close to the holographic bound, $\approx \; 1.25$: as we explained above, we had no reason to expect this. Second, it is clear that, in the phenomenological model we are using here \cite{kn:sahoo}, $\mathfrak{s}T$ does in fact increase with temperature, but it \emph{does also appear to be bounded above}, by a quantity surprisingly close to the holographic prediction.

We can summarize the results of this section by stating that \emph{the upper bound on the rate of specific complexification predicted by holography probably does correspond to a real property of strongly coupled matter}, in the case of low specific angular momenta.

Our faith in the existence of such holographic bounds thus fortified, we can now ask what happens when the boundary plasma is produced in peripheral collisions, where the specific angular momentum need \emph{not} be small.

\addtocounter{section}{1}
\section* {\large{\textsf{4. Rate of Change of the Specific Complexity: Large $\mathfrak{j}$}}}
Thus far, we have been considering the situation which corresponds, experimentally, to ``central'', or (effectively) ``head-on'' collisions of heavy ions. The STAR collaboration has also studied \cite{kn:STARcoll,kn:STARcoll2} peripheral collisions, which can generate relatively large values of the specific angular momentum. We now wish to consider, then, the possibility that $\mathfrak{j}$ is \emph{not} small.

We claimed earlier that, for all fixed $\mathfrak{j},$ $\Gamma\left(T^*L,\;\mathfrak{j}/(cL)\right)$ is a bounded monotonically increasing function of $T^*L$. For each $\mathfrak{j}$, then, the (least) upper bound on $\Gamma\left(T^*L,\;\mathfrak{j}/(cL)\right)$ as a function of $T^*L$ is given by the limit as $T^*$ tends to infinity:
\begin{equation}\label{N}
\Gamma\left(T^*L,\;\mathfrak{j}/(cL)\right)\;<\; \lim_{T^* \rightarrow \infty}\,\Gamma\left(T^*L,\;\mathfrak{j}/(cL)\right).
\end{equation}
This limit can be evaluated explicitly, and, after a lengthy algebraic manipulation, one obtains (from equation (\ref{L})) the least upper bound on the rate of specific complexification:
\begin{equation}\label{O}
{d\mathfrak{C}\over dt} \; < \; {4 c^2\over 3\hbar}\,\left(2\;-\;\sqrt{1\;+\;3 \left(\mathfrak{j}/(cL)\right)^2}\right).
\end{equation}
The function on the right being monotonically decreasing, we obtain incidentally a universal (for systems dual to AdS$_5$-Kerr black holes) upper bound on the rate of specific complexification:
\begin{equation}\label{P}
{d\mathfrak{C}\over dt} \; < \; {4 c^2\over 3\hbar} \; \approx \; 1.14 \times 10^{51}/(\m{kg}\cdot \m{s}).
\end{equation}
As we know, for ordinary matter this is not a useful condition; but the right side is in fact just above the largest value seen in Figure 2 for the QGP.

As $\mathfrak{j}/(cL)$ increases, the bound becomes stricter; but at first it does so very slowly. For example, when $\mathfrak{j}/(cL)$ reaches $0.2$, the right side of (\ref{O}) falls only very slightly, to (as we saw in the preceding Section) $1.07 \times 10^{51}/(\m{kg}\cdot \m{s})$. After that, however, the decline rapidly gains pace, as can be seen in Figure 3. This is the sense in which $\mathfrak{j}/(cL) = 0.2$ is ``small'': this value of the specific angular momentum has hardly any effect on the upper bound.

\begin{figure}[!h]
\centering
\includegraphics[width=0.6\textwidth]{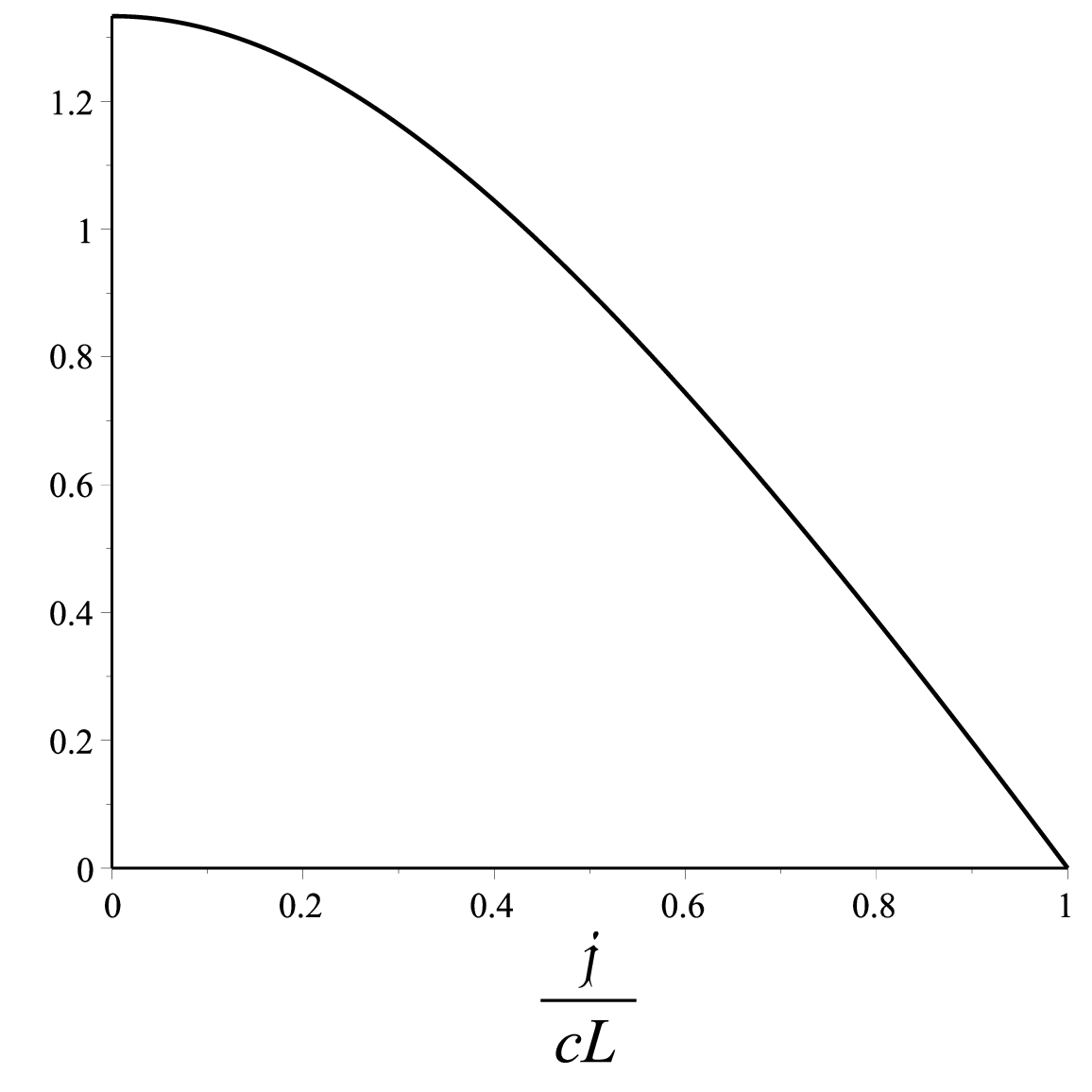}
\caption{The upper bound on the rate of growth of specific complexity, in multiples of $c^2/\hbar$, as a function of $\mathfrak{j}/(cL)$.}
\end{figure}

It is clear both from the inequality (\ref{O}) and from Figure 3 that, as $\mathfrak{j}/L$ approaches $c$, the upper bound tends to zero, and hence so does the rate itself: it seems that the specific complexification of the boundary field theory can in this manner be (almost) \emph{brought to a complete halt}. In short, \emph{if} it is possible for $\mathfrak{j}/L$ to come near to $c$, then the ``fastest complexifier'' can be slowed down to any desired degree.

Note that this is in very sharp contrast to the case of four-dimensional, asymptotically flat Kerr black holes. For while it is true that increasing the rate of rotation (at fixed temperature) has the effect of reducing the specific entropy in that case, the extent of the reduction is always limited. That is, the specific entropy is bounded \emph{below} by a strictly non-zero value (which depends inversely on the fixed temperature, but not on anything else) in that case, as is discussed in detail in \cite{kn:109}. But this is misleading: in the five-dimensional, asymptotically AdS case, \emph{there is no lower bound} to the specific entropy and therefore to the rate of specific complexification.

We evidently need to ask whether experimental or phenomenological studies of peripheral heavy ion collisions can help us to decide whether $\mathfrak{j}/L$ does in fact come close to $c$. As we discussed earlier, we take this to mean: does the ``highly vortical'' QGP described in \cite{kn:STARcoll,kn:STARcoll2} rotate so rapidly that the matter in the vortices moves at speeds close to the speed of light?

Contrary to what one might think, this may well \emph{not} be the case, because the vortices have to be formed in a medium which is extremely viscous. (The viscosity of the QGP is often said to be ``small'', but in fact it is only small relative to the entropy density: in absolute terms it is exceedingly large, in fact by far the largest known for any fluid.)

In \cite{kn:STARcoll} the reported angular velocity is $9\,\pm 1\,\times 10^{21}\,\cdot\,$s$^{-1},$ which corresponds to $0.03 \;\m{fm}^{-1}$ in natural units; this refers to average quantities computed over the volume of the plasma sample and over the observed range of impact energies. For such small systems, with diameter on the order of a femtometre, this is \emph{not} a very large figure, as is emphasised in \cite{kn:declan}: it does not lead to velocities near to that of light. It may well be, then, that even the most vortical RHIC plasmas complexify rapidly.

This leads us to consider the lead-lead collisions studied in the ALICE experiment at the LHC \cite{kn:ALICEoverview}. Here the typical impact energy is 2.76 TeV, the energy density \cite{kn:aliceenergy} is about 2.3 times larger than in the 200 GeV RHIC collisions, but the angular momentum density is computed \cite{kn:93} to be about 13.5 times larger for a given centrality. Thus it is quite possible that the maximum velocities attained in the vortices at this impact energy are a substantial fraction of the speed of light, and so it may be that the rate of specific complexification of the corresponding plasmas is significantly decreased, perhaps even to near zero. One then has to ask: how would ultra-slow complexification manifest itself observationally, if at all?

Unfortunately, it is very difficult to distinguish rapidly rotating plasmas in these experiments from their non-rotating counterparts. To understand why, note that what is actually observed in these experiments is the polarization of protons emitted as $\Lambda$/$\overline{\Lambda}$ hyperons, produced in the collision, decay. This polarization is however apparently disrupted by the higher temperatures as the impact energy increases. The upshot is that the observed effect actually becomes \emph{smaller} at higher energies, as can be seen clearly in \cite{kn:STARcoll,kn:STARcoll2}. It is already difficult to detect at 200 GeV impact energy, and it is not detectable at all in the ALICE observations \cite{kn:bed}.

Thus, while it is rather plausible that the extreme vorticities arising in these experiments lead to very low complexification rates, we have at present no way to identify the relevant collisions.

We might instead turn to phenomenological studies of the vortical QGP, analogous to \cite{kn:sahoo}; but these are as yet in their infancy (see for example \cite{kn:kshitish,kn:dutta} and their references), and the studies so far are necessarily based on very specific models. In \cite{kn:kshitish} there is, as the holographic model predicts, an increase in both the entropy and energy densities at fixed temperature, as the vorticity increases. (Recall from our discussion above that both the entropy and the physical mass of an AdS$_5$-Kerr black hole do increase, at fixed temperature, with increasing specific angular momentum; it is only the specific entropy that decreases.) However, these changes are only significant at extremely large vorticities (up to 2 fm$^{-1}$), far beyond values attained in current experiments. It will be interesting to see whether this remains the case as these models are further developed. Similar questions arise for the plasmas which existed in the early Universe \cite{kn:raf}.

To summarise: no definite evidence for conditions that would imply substantial decreases in rates of complexification with increasing QGP vorticity is seen in current experiments or in the early phenomenological models. In view of the difficulties in observing QGP vorticity at all ---$\,$ the first definitive announcement appeared in 2017 ---$\,$ this is perhaps to be expected. We believe that the effect is real, however, and we feel that there is reason to hope that progress on the experimental \cite{kn:nica} and phenomenological fronts will eventually help to settle these questions, if indeed the rate of complexification is associated with observable effects.

We can still ask a theoretical question, however: is the purported ability of large specific angular momenta (effectively) to halt the complexification of the boundary matter reflected in the interior of the bulk black hole? We now attempt to answer this.

\addtocounter{section}{1}
\section* {\large{\textsf{5. Large $\mathfrak{j}$ Slows the Interior Black Hole Dynamics}}}
The proposal of \cite{kn:suss1,kn:suss2,kn:suss3,kn:suss4,kn:poli} is that complexification in the boundary field theory is dual to the evolution of the interior of the bulk black hole. According to our findings here, this should mean that increasing the specific angular momentum of an AdS$_5$-Kerr black hole should retard the evolution of its interior. Let us see how that might work.

We should emphasise immediately that the geometry we are using is probably valid just under the event horizon of a rotating AdS$_5$ black hole, but not deep inside: see \cite{kn:ham} and its references for this. The ``collapsing wormhole'' picture at small values of $r$ is probably still valid qualitatively, but it is not described by the AdS$_5$-Kerr metric.

We will therefore proceed by confining attention to internal spatial sections which stay away from the ``late'' region, assuming that the metric given in (\ref{A}) above continues to be at least approximately valid on such sections. That is, we only consider values for $r$ smaller than but very close to $r_{\textsf{H}}.$

As usual, $r$ is timelike just under the event horizon, but it is very close to being null. If, as promised, we remain at a value of $r$ smaller than but very close to $r_{\textsf{H}},$ then this time coordinate is approximately the one used in \cite{kn:suss1,kn:suss2,kn:suss3,kn:suss4,kn:poli} (see also \cite{kn:christ}) to measure the rate of change of the interior geometry ---$\,$ except that it becomes \emph{smaller} towards the future, a confusing property which must be kept in mind throughout the discussion which follows.

Fixing such a value of $r$ means that we are considering a four-dimensional spacelike slice of the interior spacetime. We will study the evolution of the volume of a piece of this slice, taking $r$ to be our time coordinate.

We take a finite interval $\tau$ of values of $t$, which is spacelike, and let the angular coordinates range as usual ($0$ to $2\pi$ for $\phi$ and $\psi$, $0$ to $\pi/2$ for $\theta$ ---$\,$ bear in mind that these are Hopf coordinates). The volume, $\Omega,$ of this four-dimensional domain is a function of time (that is, of $r$). We propose to study the time evolution of the interior geometry by studying this function of time.

One finds that, at ``early'' times, that is, for values of $r$ slightly smaller than $r_{\textsf{H}},$ $\Omega$ increases with time (that is, with \emph{decreasing} $r$); this is in agreement with the discussions in \cite{kn:suss1,kn:suss2,kn:suss3,kn:suss4,kn:poli,kn:christ}. Eventually, however, for small enough $r$, the volume reaches a maximum and then begins to decrease, as is to be expected near to $r = 0$. As stated earlier, we stay away from this region, so, for us, the derivative of $\Omega$ with respect to $r$ is always negative (with $\Omega$ increasing in the direction of small $r$).

We stress in passing that we are not (necessarily) making a connection here with the well-known ``complexity = volume'' hypothesis \cite{kn:poli}. All we are seeking is evidence that large specific angular momenta strongly suppress the evolution of the geometry of a spacelike ``sample'' of the interior.

Using (\ref{A}) one finds, after computing the relevant determinant and simplification, that the element of volume is given by
\begin{equation}\label{Q}
\m{d}\Omega \;=\; {r\,\sqrt{|\Delta_r|}\,\rho \,\sin(\theta)\,\cos(\theta)\,\m{d}t\m{d}\theta\m{d}\phi\m{d}\psi\over \Xi}\,,
\end{equation}
and so we have, assuming $a \neq 0$,
\begin{equation}\label{R}
\Omega \;=\; {4\,\pi^2 \tau r\,\sqrt{|\Delta_r|}\,\left[\left(a^2 + r^2\right)^{3/2}\,-\,r^3\right]\over 3 a^2\,\Xi}.
\end{equation}
Here we can express $|\Delta_r|$ in the following way: eliminating $M$ using equation (\ref{I}), we have
\begin{equation}\label{S}
|\Delta_r|\;=\; \left(r_{\textsf{H}}^2+a^2\right)\left(1 + {r_{\textsf{H}}^2\over L^2}\right)\;-\; \left(r^2+a^2\right)\left(1 + {r^2\over L^2}\right).
\end{equation}

As is well known, volumes inside black holes can be extremely large, including in the rotating case: see for example \cite{kn:xiaoyen} and its references. The time rate of change might therefore be large for no other reason than that $\Omega$ itself is large ---$\,$ for example, \emph{both} $\Omega$ and (the magnitude of) its derivative with respect to $r$ can be made large simply by choosing $\tau$ to be sufficiently large, or $a$ to be sufficiently close to $L$. To avoid being misled in this way, we will focus on the logarithmic derivative of $\Omega$ with respect to $r$, which we denote by $\lambda(\Omega)$.

As always, we use (\ref{G}) to express $a$ in terms of $\mathfrak{j},$ and (\ref{H}) to express $r_{\textsf{H}}$ in terms of $T$ and $\mathfrak{j}$, so that now $\Omega$ and $\lambda(\Omega)$ can be regarded as functions of $T$ and $\mathfrak{j}$. As usual, they are both (particularly $\lambda(\Omega)$) very complicated functions, and a numerical investigation is required to produce comprehensible results.

We begin by recalling \cite{kn:109} that, for fixed temperature $T$, $r_{\textsf{H}}$ (for the case of the ``large'' black hole) is an increasing function of $\mathfrak{j}.$ The smallest possible value of $r_{\textsf{H}}$ here therefore occurs for the ``large'' AdS$_5$-Schwarzschild black hole, for which it is easily shown to be given by
\begin{equation}\label{T}
r_{\textsf{H}}^{\textsf{AdS}_5\textsf{Large Sch}}\;=\;{L^2\over 2}\,\left(\pi T^* + \sqrt{\pi^2 T^{*2} - (2/L^2)}\right).
\end{equation}
(where again we use $T^* = k_{\textsf{B}}T/(\hbar c)$ for convenience). The values of $r$ of interest to us here are therefore a little smaller than this. As $\mathfrak{j}$ increases, so does $r_{\textsf{H}}$; it tends to $\pi T^* L^2$ as $\mathfrak{j}/L$ nears $c$.

Let us consider the case $T^*L = 1$ for example. Then the right side of (\ref{T}) is approximately $2.97344\,L$; hence, if we choose $r$ smaller than this, it will be smaller than $r_{\textsf{H}}$ for all values of $\mathfrak{j}$. (As $\mathfrak{j}/L$ increases from zero to $c$, $r_{\textsf{H}}$ rises from this value towards $\pi L$.)

If we choose $r = 2.973\,L,$ then we are considering points just below the event horizon when $\mathfrak{j}$ is zero. As $\mathfrak{j}$ increases, the event horizon moves out, but not very far (to at most $3.14 \,L$), so we are still near to the event horizon, in this sense, for all values of $\mathfrak{j}.$

The graph of $|\lambda(\Omega)|$ (multiplied by $L$ so that it is dimensionless) in this case is shown ---$\,$ we take the absolute value for reasons explained above ---$\,$ in Figure 4.
\begin{figure}[!h]
\centering
\includegraphics[width=0.95\textwidth]{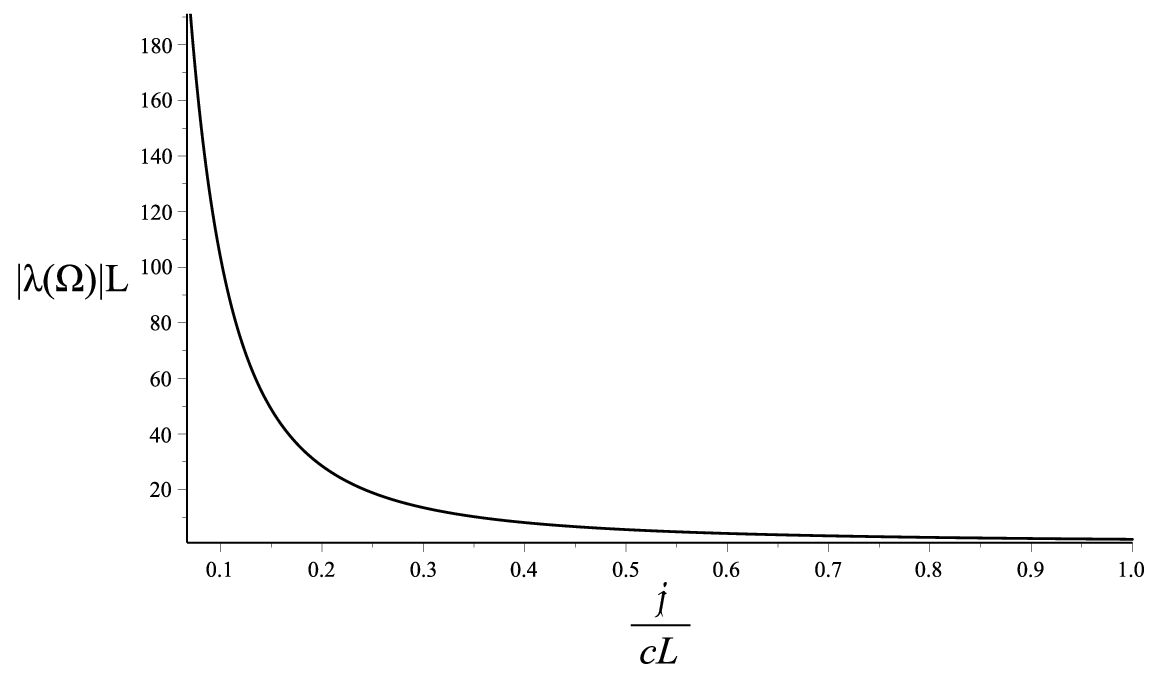}
\caption{$|\lambda(\Omega)|L$ for $T^*L = 1, r = 2.973 L$, as a function of $\mathfrak{j}/(cL)$.}
\end{figure}
We see that values of $\mathfrak{j}/L$ near to $c$ do indeed drastically slow down the evolution of the volume of this region of the interior. In fact, the difference is greater than can be shown legibly in the diagram: the actual value of $|\lambda(\Omega)|L$ at $\mathfrak{j} = 0$ is approximately 1100, while on the other hand the limiting value as $\mathfrak{j}/L$ tends to $c$ is around 2.1. Thus, while the evolution of the interior is not literally brought to a halt by high specific angular momenta, the reduction of its rate of change by a factor of over 500 is certainly striking. If $r$ is taken to be closer to $r_{\textsf{H}}$, then this factor rapidly becomes still larger, with no limit we have been able to detect numerically.

A reasonable summary is to state that there is a definite correspondence between the effects of high specific angular momenta on the complexification of the boundary matter and on the evolution of the bulk black hole interior: in both cases, there is a strong slowing effect.

\addtocounter{section}{1}
\section* {\large{\textsf{6. Conclusion: Slow Complexification and Traversability}}}
In peripheral heavy-ion collisions, large magnetic fields \cite{kn:hao} are generated, in a manner similar to the classical \emph{Barnett effect} \cite{kn:barn}. This constricts the corresponding phase space, reducing the entropy per particle at a given temperature. Detailed phenomenological studies \cite{kn:hof} confirm this for QCD at relatively low temperatures, and similar effects are predicted for the QGP \cite{kn:bali}.

Large vorticities in the QGP can be expected to have the same effect, reducing the number of degrees of freedom per particle available to ``execute gates'' at a given temperature. This admittedly somewhat vague observation makes our principal finding in this work, that (according to holography) large specific angular momenta tend to suppress the rate of complexification of strongly coupled matter, rather more understandable.

The fact that there is no limit to the extent of the predicted reduction is more surprising; it is not something for which we are prepared by a study of the four-dimensional, asymptotically flat case (where there is a strictly positive lower bound on the specific entropy, at any given non-zero temperature) \cite{kn:109}. Nevertheless, we have found confirmation of this prediction by studying the effect of large specific angular momenta on the interior dynamics of the bulk black hole: we have found that the interior evolution can be slowed by a very large factor, and we find no evidence for a bound on this factor.

To put this last observation another way: for wormholes which avoid the region of small values of $r$, high specific angular momenta can strongly retard the characteristic collapse, bringing it almost to a halt. It is natural to ask whether this effect might be so strong as to allow the wormhole to be traversable.

The answer is no under the conditions we have assumed here, for our black hole is still strictly classical and the null energy condition is satisfied everywhere in the spacetime. However, the phenomenon of slow wormhole collapse strongly suggests that rapidly rotating wormholes might very easily \emph{become} traversable after a suitable quantum perturbation, as in \cite{kn:aronwall,kn:juan}.

To put it another way, when quantum effects are taken into account, it may be that \emph{rotationally induced slow complexification of strongly coupled matter has a holographic dual description in terms of a traversable wormhole}.

In this work, we have focused our attention on slowing complexification by using large specific angular momenta to reduce the number of degrees of freedom involved in executing gates at a fixed temperature. There is a more direct way of accomplishing this: we could try to elongate the time scale for executing the gates. Since this time scale is given (as above) by $\hbar/\left(k_BT\right)$, this means using rotation to lower the temperature ---$\,$ in other words, it means that we consider rotationally near-extremal black holes. These objects, and the dual matter, complexify slowly. Following our logic above, this suggests that near-extremal AdS$_5$-Kerr wormholes might be quantum-perturbed into traversability.

We have stressed that the black holes studied in this work are \emph{not} near-extremal. We have deliberately avoided the near-extremal case for two reasons. Firstly and most obviously, real strongly coupled matter is not cold, except possibly when the baryonic chemical potential is extremely large, which is not the case for the QGP. Secondly, as we mentioned earlier, it seems likely that near-extremal black holes (with spherical event horizon topology) are pathological to some extent \cite{kn:horo1,kn:horo2,kn:108,kn:horo3,kn:tur}.

Nevertheless, the near-extremal case has received a great deal of attention, so one can ask whether any evidence has been found to confirm our conjecture that nearness to (rotational) extremality might be associated with traversable wormholes. In fact, precisely this association has recently been established: it was found in \cite{kn:bilotta} that certain near-extremal rotating black holes, after the back-reaction effect discussed in \cite{kn:aronwall,kn:juan} is taken into account, do in fact contain traversable wormholes.

The black holes considered in \cite{kn:bilotta} are four-dimensional and asymptotically flat, and so they are much simpler than the black holes we have studied here. Nevertheless, it is reasonable to hope that the methods of \cite{kn:bilotta} can be extended to near-extremal AdS$_5$-Kerr black holes, and ultimately also to strictly \emph{non-extremal} AdS$_5$-Kerr black holes. One might then be able to prove that quantum perturbations of the slowly collapsing wormhole we studied in the previous Section do in fact render it traversable.

In conclusion, let us discuss the most basic question in this subject, namely: \emph{what is the physical meaning} of (specific) complexity, in real, physical systems such as the QGP? The reason we have been able to make some progress in this work is simply the fact that $\mathfrak{s}\,T$, the \emph{rate} of specific complexity growth, \emph{is} clearly a physical quantity. But what is the physical relevance of the underlying complexity itself?

About this we can, for the moment, only speculate. We offer the following suggestions.

It seems clear that specific complexity cannot be ``measured'' directly; but perhaps it might have some indirect effect which can.

The quantum circuit complexity of a state is the minimal number of (2-qubit) gates needed to prepare that state. Naively, it is a measure of how ``difficult'' it is to prepare the state. But there is another, though closely related, way of thinking about it: complexity is a measure of how hard it is to exert effective \emph{control} over the state: intuitively, we might say that a state which requires many gates to prepare is one that is hard to control in a precise way. This is consistent with, for example, the intuition that chaotic systems generically have high complexity (see for example \cite{kn:joanna}).

If the state is one which, for some purpose, needs to be controlled precisely, then a high degree of complexity might make it difficult or impossible to use that state for that purpose.

According to \cite{kn:aronwall,kn:juan}, the holographic dual of the formation of traversable wormholes in the bulk black hole is the existence of some system on the boundary which implements \emph{quantum teleportation} \cite{kn:telep}. Now it is notoriously difficult to construct systems implementing teleportation: it can only be done when the system can be manipulated very precisely. We therefore speculate that teleportation requires that the boundary system should \emph{not} have a high complexity.

Of course, the most straightforward way to prevent complexity from becoming large is to reduce its rate of growth. The tentative picture we therefore arrive at is the following: large specific angular momenta in the bulk black hole are associated with a slowdown in the collapse of the wormhole, allowing it to be (quantum) traversable; the dual description is that large specific angular momenta on the boundary slow down complexification and prevent it from hampering the establishment of quantum teleportation in strongly coupled matter.

The possible existence of quantum teleportation in the vortical QGP does not appear to have been investigated as yet, but it might well have important or even drastic consequences. However, this is not entirely unexpected: recent work on the ``ultra-vortical'' QGP \cite{kn:buzztuch} does suggest that it might well differ qualitatively from its more slowly rotating counterparts, when the angular velocity, $\omega$, is so large that $1/\omega$ is smaller than the mean free path in the plasma. Perhaps this is the regime in which quantum teleportation plays some as-yet unknown role.

In any case, this speculative discussion shows that a physical role for complexity is not out of the question.

\addtocounter{section}{1}
\section*{\large{\textsf{Acknowledgement}}}
The author is grateful to Professors Gao Sijie and Ong Yen Chin, and to Dr. Soon Wanmei, for useful comments.

\end{document}